\documentclass[11pt,paper,letterpaper]{JHEP3}
\usepackage{epsfig}
\usepackage{psfrag}
\usepackage{amsmath}
\newcommand{\bq}{\begin{eqnarray}}
\newcommand{\eq}{\end{eqnarray}}
\newcommand{\bqs}{\begin{eqnarray*}}
\newcommand{\eqs}{\end{eqnarray*}}

\def\pd{\partial}

\def\g{\gamma}

\def\s{\sigma}

\def\S{\Sigma}
\title{The Petrov type of the five-dimensional Myers-Perry metric}
\author{Pieter-Jan De Smet\\
C.~N. Yang Institute of Theoretical Physics\\
State University of New York\\
Stony Brook, NY 11794-3840, USA\\
E-mail: {\tt Pieterj@insti.physics.sunysb.edu} }
\preprint{YITP-SB-03-63 \\{\tt gr-qc/0312021}}
\abstract{We point out that the Myers-Perry metric in five dimensions is algebraically
special. It has Petrov type~\underline{22}, which is the Petrov type of the 
five-dimensional Schwarzschild metric.}
\begin{document}
\section{Introduction}
In this article, we calculate the Petrov type of the five-dimensional Myers-Perry
metric~\cite{Myers-Perry}. 
The five-dimensional Myers-Perry (MP) metric is the generalization to five 
dimensions of the four-dimensional Kerr-metric.  
The Kerr-metric, which describes the gravitational field of a rotating star,
and its static limit, the Schwarzschild metric, have both Petrov type~$D$. 
It is remarkable to see that, although the Kerr-metric has fewer symmetries
than the Schwarzschild metric, it does still have the same Petrov type.

We show that the same holds in five dimensions. The 
five-dimensional MP metric, which describes a rotating black hole
in five dimensions, and its static limit, the five-dimensional Schwarzschild
metric, have both Petrov type~\underline{22}. Again, we see that although
the five-dimensional MP metric has fewer isometries than the
five-dimensional Schwarzschild metric, it does have the same Petrov type.
The remainder of the article is organized as follows.
In Section~2, we give a review of the 
five-dimensional Petrov classification. The Petrov type of the MP metric 
is given in Section~3. We conclude in Section~4.
\section{Review of the five-dimensional
Petrov classification}\label{s:Petrov}
We only give a brief review of this classification, a longer discussion can 
be found in ref.~\cite{0206106}. We need to introduce two objects, the 
\textit{Weyl spinor} and the \textit{Weyl polynomial}. 
The Weyl spinor $\Psi_{abcd}$
is the spinorial translation of the Weyl tensor $C_{ijkl}$
$$\Psi_{abcd} = (\g_{ij})_{ab} (\g_{kl})_{cd}C^{ijkl}.$$
Here, $\g_{ij} = \frac{1}{2} [\g_i,\g_j ]$, where $\g_i$ are
the $\g$-matrices in five dimensions. 
In this article, we use the following representation
$\g_1 = i \s_1 \otimes 1$, $\g_2 = \s_2\otimes 1$, $\g_3 = \s_3\otimes \s_1$,
$\g_4 = \s_3 \otimes \s_2$ and $\g_5 = \s_3\otimes \s_3$. The Weyl spinor is 
symmetric in all its indices.
The Weyl polynomial $\Psi$ is a homogeneous polynomial of degree four in four variables:
$$
\Psi = \Psi_{abcd} x^a x^b x^c x^d\ .
$$
The Petrov type of a given Weyl tensor is the number and
multiplicity of the irreducible
factors of its corresponding Weyl polynomial $\Psi$. 
In this way, we obtain 12 different
Petrov types, which are depicted in figure~\ref{fig:Ptypes}.
\FIGURE[ht]{
\begin{psfrags}
\psfrag{1}[][]{4}
\psfrag{2}[][]{31}
\psfrag{3}[][]{22}
\psfrag{4}[][]{211}
\psfrag{5}[][]{\underline{22}}
\psfrag{6}[][]{2\underline{11}}
\psfrag{7}[][]{1111}
\psfrag{8}[][]{\underline{11} \underline{11}}
\psfrag{9}[][]{11\underline{11}}
\psfrag{10}[][]{$\Psi = 0$}
\psfrag{11}[][]{\underline{1111}}
\psfrag{12}[][]{1\underline{111}}
\epsfig{file=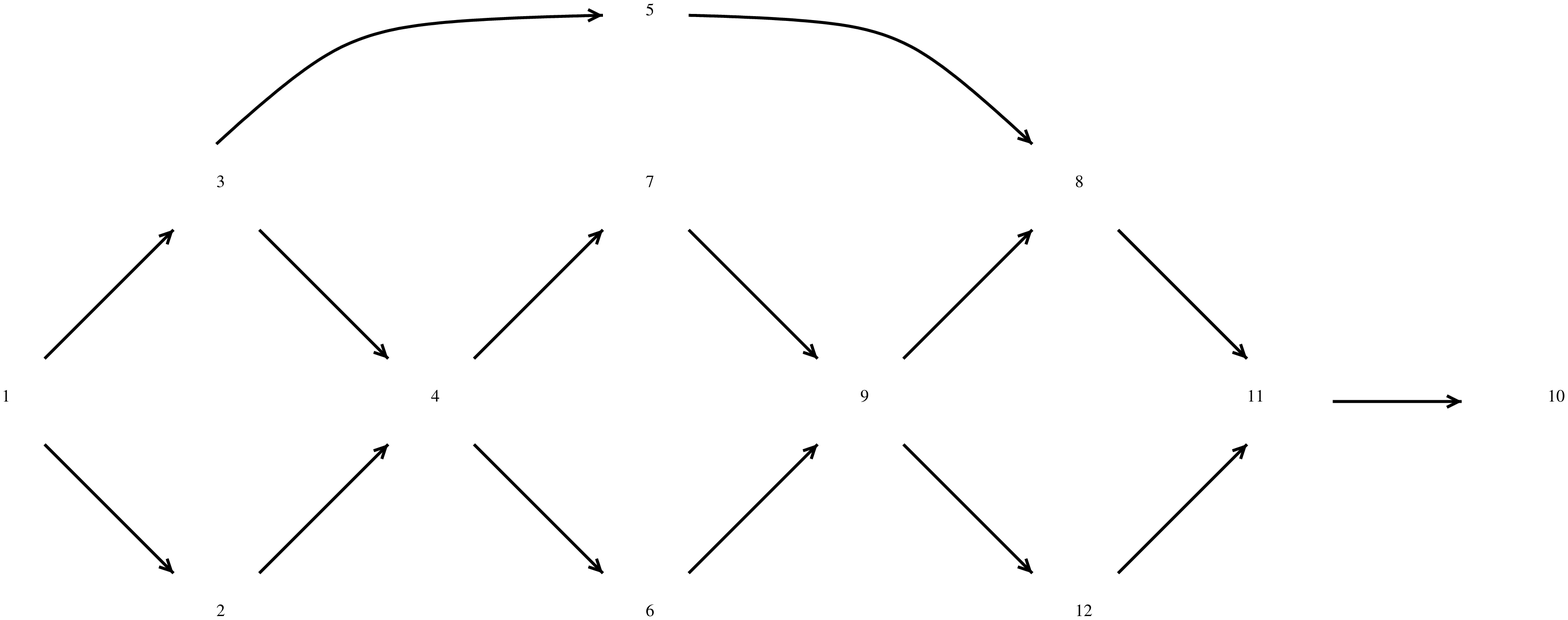,width=10cm,height=5cm}
\end{psfrags} 
\caption{The 12 different Petrov types in 5 dimensions.}
\label{fig:Ptypes}
}
We use the following notation. The number denotes the degree of the
irreducible factors and underbars denote the multiplicities. For example, a Weyl
polynomial which can be factorized into two different factors, each having
degree~2, has Petrov type~22. If the two factors of degree~2 are the same, the 
Petrov type is~\underline{22}.
\section{The Myers-Perry metric has Petrov type~\underline{22}}
The five-dimensional rotating black hole is described by the Myers-Perry metric~\cite{Myers-Perry}
\begin{equation*}
\begin{split}              
ds^2=& -dt^2 + \frac{2m}{\rho^2} \left[ dt - a \sin^2 \theta d\phi
- b \cos^2\theta d\psi \right]^2 \\
&+ \frac{\rho^2}{R^2} dr^2 + \rho^2 d\theta^2  + \S^2_a \sin^2\theta d\phi^2 + 
\S^2_b \cos^2\theta d\psi^2,
\end{split}
\end{equation*}
where $\rho^2 = r^2 + a^2 \cos^2\theta + b^2 \sin^2\theta$,
$\S^2_a = r^2 + a^2$, $\S^2_b = r^2 + b^2$ and 
\begin{equation*}
\begin{split}
R^2 &= \left[ \S^2_a \S^2_b - 2 m r^2\right]/r^2.\\
\end{split}
\end{equation*}
We choose the following tetrad
\begin{equation*}
\begin{split}
e_1 =& \frac{1}{r^2 R \rho} \left[ \S^2_a \S^2_b \pd_t + a \S^2_b \pd_\phi + 
b \S^2_a \pd_\psi\right]\\
e_2 = &\frac{1}{r \rho \sin\theta}\left[ \S_b 
\left( \pd_\phi + a \sin^2\theta \pd_t \right) + b \S_a V \right]\\
e_3 = &\frac{1}{r \rho \cos\theta}\left[ \S_a 
\left( \pd_\psi + b \cos^2\theta \pd_t \right) -a \S_b V \right]\\
e_4 =& \frac{R}{\rho} \pd_r\\
e_5=& \frac{1}{\rho} \pd_\theta.\\
\end{split}
\end{equation*}
Here, we have used the vectorfield 
$$V = \frac{1}{\S_a \S_b + r \rho}\left( a \sin^2\theta \pd_\psi 
- b \cos^2\theta \pd_\phi\right).$$
A straigthforward calculation gives the Weyl polynomial
$$\Psi = - \frac{48 m r^2}{\rho^6}
\left[x^2 - y^2 + z^2 - t^2 - 2 f ( x y + z t ) - 2 i g (x z + y t) \right]^2,$$
where
\begin{equation*}
\begin{split}
f &= \frac{b \sin\theta}{r} \frac{ a^2 \S_b \rho + (b^2 - a^2) r \S_a \cos^2
\theta}{\S_a^2 \S_b^2 - r^2 \rho^2},\\
g &= \frac{a \cos\theta}{r} \frac{ b^2 \S_a \rho + (a^2 - b^2) r \S_b \sin^2
\theta}{\S_a^2 \S_b^2 - r^2 \rho^2}.\\
\end{split}
\end{equation*}

This polynomial is the square of a polynomial of degree~2. Therefore, the MP
metric has Petrov type~\underline{22}.
\section{Conclusions and topics for further research}
In this article, we showed that the five-dimensional MP metric has the same Petrov
type as its static limit, namely Petrov type~\underline{22}. Some open problems 
are the following.
\begin{itemize}
\item
Recently, Emparan and Reall found a black rotating ring~\cite{ring}, see ref.~\cite{Teo} for 
easier coordinates. It would be nice to know its Petrov type.
\item
In four dimensions, adding electric charge to the rotating star does not change its
Petrov type; the Kerr-Newmann metric has Petrov type~$D$. In five dimensions, 
the story is more complicated. The metric of an electrically charged rotating black
hole is only known in five dimensions when there is a specific Chern-Simons term in the 
action. The particular form of this Chern-Simons term is dictated by supersymmetry. The 
charged rotating black hole in this theory is described by the BMPV metric~\cite{BMPV}, which
was found by using duality. It would be good to calculate its Petrov type. The charged
rotating black hole is not known when this Chern-Simons term has an arbitrary (or even zero)
coefficient. It remains to be seen if it can be found within the class of algebraically
special metrics.
\end{itemize}
\section*{Acknowledgments}
This work has been supported in part by the NSF grant PHY-0098527.

\end{document}